\documentclass[proof]{WileyASNA-v2}

\articletype{Article Type}%

\received{}
\revised{}
\accepted{}

\newcommand{\xmm}{{\it XMM-Newton}}

\begin{document}

\title{XMM-Newton Publications from 2000-2024\protect\thanks{*Corresponding author: Jan.Uwe.Ness@esa.int}}

\author[1]{Jan-Uwe Ness*}

\author[1]{Norbert Schartel}

\author[1]{Maria Santos-Lleo}

\authormark{J.-U.Ness \textsc{et al}}

\address[1]{\orgdiv{European Space Astronomy Centre (ESAC)}, \orgname{European Space Agency}, \orgaddress{Camino Bajo del Castillo s/n, 28692 Villanueva de la Ca\~nada\state{Madrid}, \country{Spain}}}

\corres{*Corresponding author name, This is sample corresponding address. \email{jan.uwe.ness@esa.int}}

\abstract{\xmm\ is an ESA space X-ray observatory launched on 11 December 1999, and after 25 years,
a study is presented demonstrating that the data of the mission are efficiently used by an engaged and
productive community.
The total number of refereed papers published between 2000 and 2024 is 8486. These papers have a total
of 15627 different authors, including 3292 unique first authors. The total available science time during
this period amounts to 556\,Ms (16894 observations) 87\% of which (84\% of observations) have been used
in at least one refereed publication, excluding primary catalogue papers. Accounting
for multiple use, the observation time has been over-used by a factor of up to 15 in dedicated publications
and even a factor of up to 30 when including a small fraction of papers classified as survey/catalogue.
The speed with which observations are published for the first time peaks around 2 years and is thus longer than
the proprietary period for Guest-Observer (GO) observations. A strong secondary peak at 3 years suggests
that data not published by the proposing teams are picked up by the community, then also taking about 2 years to be published.
The publication rate remains stable at $\sim400$ refereed articles per year with a small increase in recent
years, owed to an increased use of primary \xmm\ catalogues.
95\% of articles focus on specific data using $<49$ observations while 99\% of articles use less than 239
observations.
Since 2010, the annual number of first-time authors has remained relatively constant, equalling the number of last-time authors. This implies that the number of scientists engaged in research utilising XMM-Newton data has remained constant at 4300, of whom 570 are lead (first) authors.
A histogram of the activity period demonstrates that 51\% of first authors publish for only one year, 24\% of first authors are active for up to six years, and 25\% were active for more than six years. We further identify a highly productive core community of approximately 120 scientists publishing an \xmm\ article at least every two year as $1^{st}$authors.
When all authors are considered, the figures are 40\%, 25\% and 35\%, respectively. The considerable number of time-limited activities may 
reflect a high level of utilisation during the early stages of a research career, from Master’s studies through the PhD and initial postdoctoral years.
The trends assessed indicate a vital community with positive perspectives and a continued, active interest in XMM-Newton for the future.
}


\maketitle

\section{Introduction}\label{sec1}

After 25 years in orbit, \xmm\ continues to operate as ESA's flagship mission. In 2014, the productivity
of \xmm\ was presented by \cite{xmmpubs} for the first 12 years of operations, based on 3272 publications making direct use of \xmm\ data.
More recently, \cite{guido2024} compared \xmm\ publications with those of other ESA missions and analysed the nationalities of the authors. In this work we present publication statistics up to the end of 2024 focusing on the research community and data use.
Since the last report of publication statistics in 2014, the number of refereed publications has grown to 8486,
with 7500 of them making direct use \xmm\ data.
In \S\ref{sect:meth}, we describe how we identified publications, including classifications beyond
direct use of data. We further describe how we identified the observations used in the respective publications.\\

The results are presented in \S\ref{sect:results} with a focus on author
statistics in \S\ref{sect:pubsstat} aiming to characterise the \xmm\ community in terms of size and activity
and in \S\ref{sect:usestat} the usage of \xmm\ observations.

\section{Methods}\label{sect:meth}

The generation of the \xmm\ publication list and determination of Observation Identifiers (ObsID) that uniquely identify observations used in articles is described in detail in \cite{xmmpubs}. Candidate articles are extracted
from the SAO/NASA Astrophysics Data System (ADS) database via its Application Programming Interface (API)
by making queries on {\it XMM} as a full-text search for each month of publication. The articles found in this way
were inspected manually to classify them into in total five classes. For these studies, 8486 articles (including 166
PhD theses) in the following classes are included:
\begin{itemize}
\item {\bf Class 1}: making use of data from \xmm\ observations or uses pipeline products,
\item {\bf Class 2}: presenting catalogues based on \xmm\ observations,
\item {\bf Class 3}: predicting numerical results for {\xmm} observations, e.g. performing simulations considering the detector response function,
\item {\bf Class 4}: describing \xmm, its instruments, operations, software, calibration or scientific impact, including the papers resulting from the "XMM-Newton: The Next Decade" symposia,
\item {\bf Class 5}: making use of primary catalogues: \xmm\ Serendipitous Source Catalogue \citep{2003AN....324...89W,2009A&A...493..339W,2016A&A...590A...1R,2019A&A...624A..77T,2020A&A...641A.136W,2020A&A...641A.137T}, \xmm\ Slew Survey Source Catalogue \citep{2008A&A...480..611S}, \xmm\ OM Serendipitous Ultra-violet Source Survey Catalogue \cite{2012MNRAS.426..903P,2017MNRAS.466.1061P}, Catalog of Serendipitous Sources Detected with the \xmm\ OM \citep{2008PASP..120..740K}.
\end{itemize}

\begin{figure}[!ht]
\resizebox{\hsize}{!}{\includegraphics{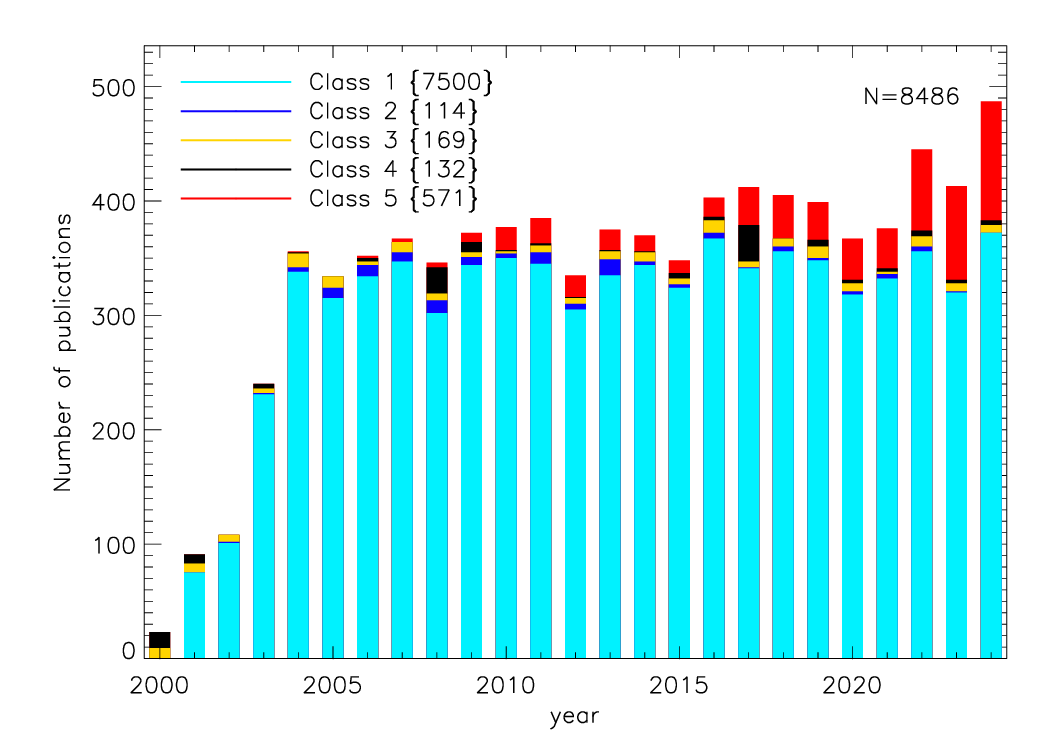}}
\caption{\label{fig:pclass}\small Number of 8486 refereed \xmm\ publications per year with classes indicating direct use of \xmm\ data (Class 1), Catalogue papers (Class 2), Prediction of \xmm\ results (e.g. simulations, Class 3), Descriptions of \xmm\ (Class 4), and articles using primary \xmm\ catalogues (Class 5). The numbers in curly brackets in the legend represent the total number of articles
in the respective classes over the entire reference period 2000-2024.
}
\end{figure}

For reproduction of results, the list {\it XMM\_pubdb2024.txt} of all 8486 publications is included in the electronic version of this article listing, in ASCII format, the bibcode, a flag indicating how the ObsID were obtained, a list of instruments used, a comma-separated list of ObsID, and the class.
The average publication rate over the 25 year reference period thus corresponds to 340 articles per year, and
a graphical representation of the evolution per year of each class is shown in Fig.~\ref{fig:pclass} where in
curly brackets in the legend, the respective numbers of articles published in the reference period 2000-2024 are given.
It can be seen that since $\sim 2016$, the publication rate is stable at $\sim 400$ articles each year, more than
one per day! By far the largest number of publications make direct use of \xmm\ data
and products while we also note the fraction of articles making use of the primary catalogues (Class 5) has increased
substantially, especially in the last three years. This trend is likely going to increase in the future as
volume and quality of the catalogues continue to increase.\\

We extracted author names from the ADS for extensive analysis of the \xmm\ community.
The names of authors are not always spelled consistently, and to minimise the effects of
counting same authors multiple times as different authors,
we removed all accents, converted special characters such as umlauts etc to
standard ASCII (e.g. \"a to ae). Further we replaced any fully spelt-out first names
by initials, and used only the respective first initials, even though this could treat
as same authors those with same last name and first names starting with the same letter,
e.g. John Smith and Jim Smith will be treated as the same person. We consider the uncertainty
introduced by this measure as smaller than treating John Smith and J. Smith as different
authors. These measures lead to some overestimates of the number
of articles per author and underestimates of the number of different authors.\\ 

\begin{figure}[!ht]
\resizebox{\hsize}{!}{\includegraphics{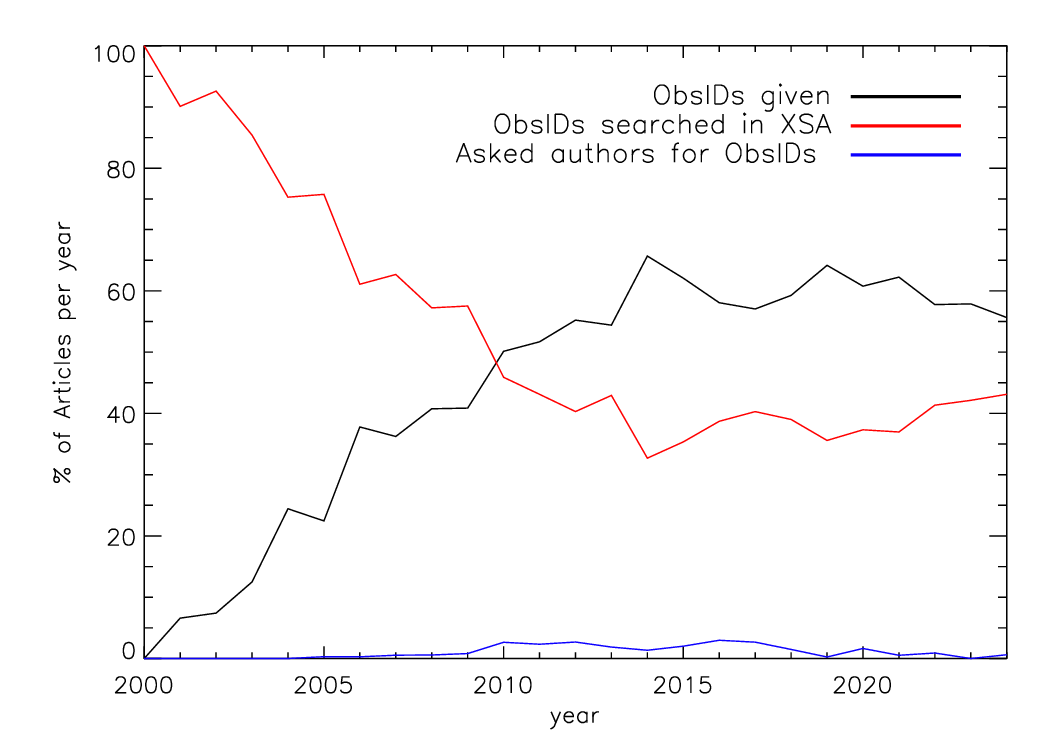}}
\caption{\label{fig:ogiven}\small Fraction of articles in which the observations used were given in terms of ObsIDs in the article text, appendix, electronic materials etc (black), where they had to be searched for in the XSA (red), and where authors provided a list of ObsIDs (blue) on request.
}
\end{figure}

The observations used in each article were extracted via manual inspection,
supported by an initial script. We screened each article for quotations of any of 16894
Observation Identifiers (ObsIDs) that uniquely associate science observations.
4190 articles contain the ObsIDs in the text, observation log tables, or electronic
appendices.
In some rare cases, false positives that the automatic script returned were removed,
e.g. from \xmm\ ObsIDs being sub-strings of the (longer) NuSTAR ObsIDs.
Also, in some cases additional ObsIDs were extracted,
e.g., in case the main article only contained a partial observation log
with full tables downloadable from the electronic version. If no ObsIDs are given,
information such as target names, observation dates etc were used to estimate
which ObsIDs were used by searching for corresponding matches in the XMM-Newton
Science Archive (XSA). For 101 articles the identification of ObsIDs was supported by the authors on request
(blue curve in Fig.~\ref{fig:ogiven}).
We excluded 57 articles from the study when the associated ObsIDs were deemed highly
uncertain.\\

The fraction of articles containing the ObsIDs has grown from initially 0\% to 
$\sim 60$\% in 2015 and remained stable since (see Fig.~\ref{fig:ogiven}).
For articles published early in the mission, the
target name is usually sufficient to uniquely identify the ObsIDs while later
in the mission, additional information such as observation dates are needed
to identify ObsIDs of sources that have been observed multiple times. Probably
for this reason, the fraction of articles indicating the ObsIDs has increased.
The \xmm\ Publication Guidelines instruct authors to include Target Name, the
Date of the Observation and the Obs-Id to ensure that the data can be uniquely identified.
Further to that, the \xmm\ project has not taken any particular measure, however,
other observatories such as {\it Chandra} (Pat Slane, Priv. Communication) have taken
strong measure to educate their authors to provide unique observation identifiers
in their articles. Since there is large overlap of authors working on {\it Chandra}
and \xmm\ data, this may have had an effect also on \xmm\ articles.\\

\section{Results}
\label{sect:results}

We present in \S\ref{sect:pubsstat} the results of publication statistics,
extracted from ADS. The uncertainties of these statistics depend on the
accuracy of the information in the ADS and completeness of our database.

In \S\ref{sect:usestat}, we present the usage statistics, depending in
addition on information extracted from the XSA. We estimate the uncertainties
of usage statistics higher than that of publication statistics:
While the accuracy of the information in the XSA will be high, the
identification of ObsIDs has not always been straight forward yielding the
largest source of uncertainty.

\subsection{Publication Statistics}
\label{sect:pubsstat}

\subsubsection{Author mentioning of \xmm}

\begin{figure}[!ht]
\resizebox{\hsize}{!}{\includegraphics{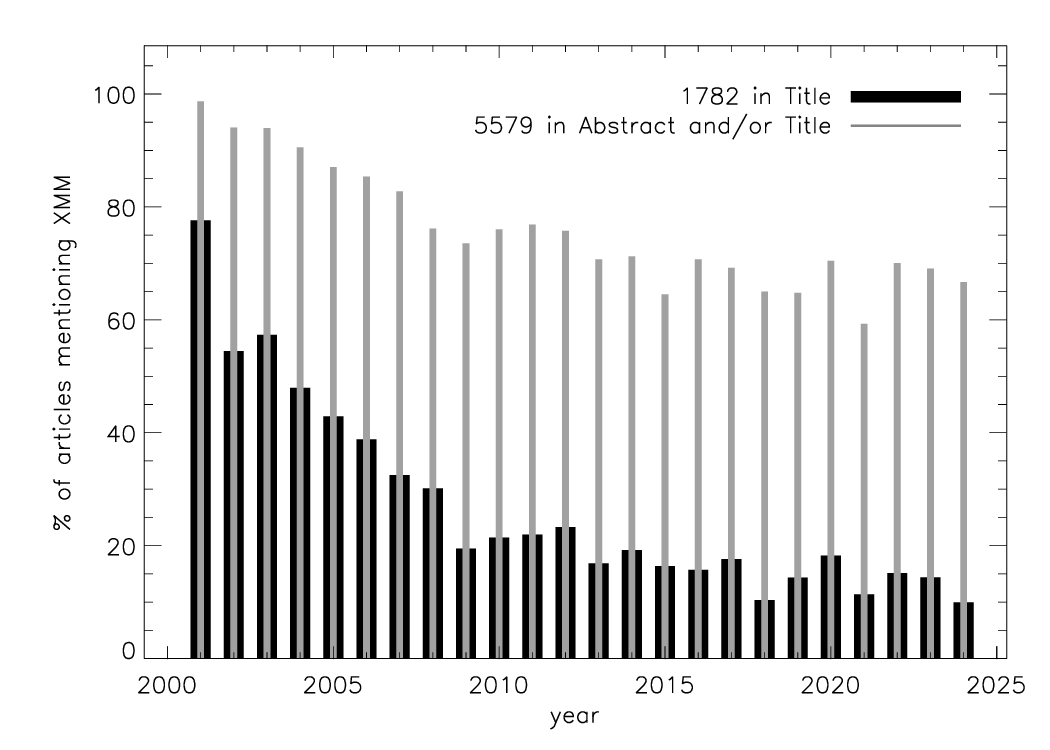}}
\caption{\label{fig:abstit}\small Evolution of the fraction of Class I articles (excluding catalogue papers), mentioning {\it XMM} in the title (black) and in abstract and/or title (grey).
}
\end{figure}

In order to assess the role \xmm\ data play in the publications, we determined the
fractions of articles when \xmm\ is mentioned in the title and abstract.
We define a measure for \xmm\ having been {\em essential} for the conclusions if
{\it XMM} is mentioned in the title (regardless of whether also mentioned in the abstract)
while we define a measure for \xmm\ having been
{\em relevant} for the conclusions if {\it XMM} is mentioned in either abstract or
title.
We consider this metric as a lower limit of relevance
because not all articles in which the conclusions depend critically on \xmm\
mention the name of the mission in the title. From the perspective of
authors, the name of the mission may be seen less relevant than the scientific
impact of the results. Especially when articles target a more general audience
such as publications in Nature or Science, authors and journal editors may
not consider it appropriate to highlight the names of observing facilities in
the title. Mention of the mission name in title/abstract can thus be seen as a necessary
but not sufficient measure for the relevance of the mission data for the conclusions.\\

The evolution of these two measures are shown in Fig.~\ref{fig:abstit}.
Articles in which \xmm\ data are {\em relevant} for the conclusions
(thus mentioned in either title or abstract) is at all times well above 60\% whereas early in
the mission, until $\sim2008$, this fraction slowly decreased from 100\% to 80\%.\\

Meanwhile, until $\sim2008$, \xmm\ was {\em essential} for
the conclusions in at least 30-60\% of all articles while after 2008, this fraction
stabilised at $\sim20$\%.\\

{\it XMM} was mentioned more often throughout the full text of the article in the
earliest years and steadily decreased in frequency until 2004, whereupon mentions
of the observatory in relevant articles has remained steady. The assumption is that
further explanation was needed when XMM-Newton was a new X-ray observatory.

\subsubsection{Author Statistics}

The total number of different authors of all 8486 \xmm\ articles is 17564 which is
reduced to 15627 after applying the streamlining of author names described above,
of which 3292 different scientists have published at least one \xmm\ article as $1^{st}$author.\\

\begin{figure}[!ht]
\resizebox{\hsize}{!}{\includegraphics{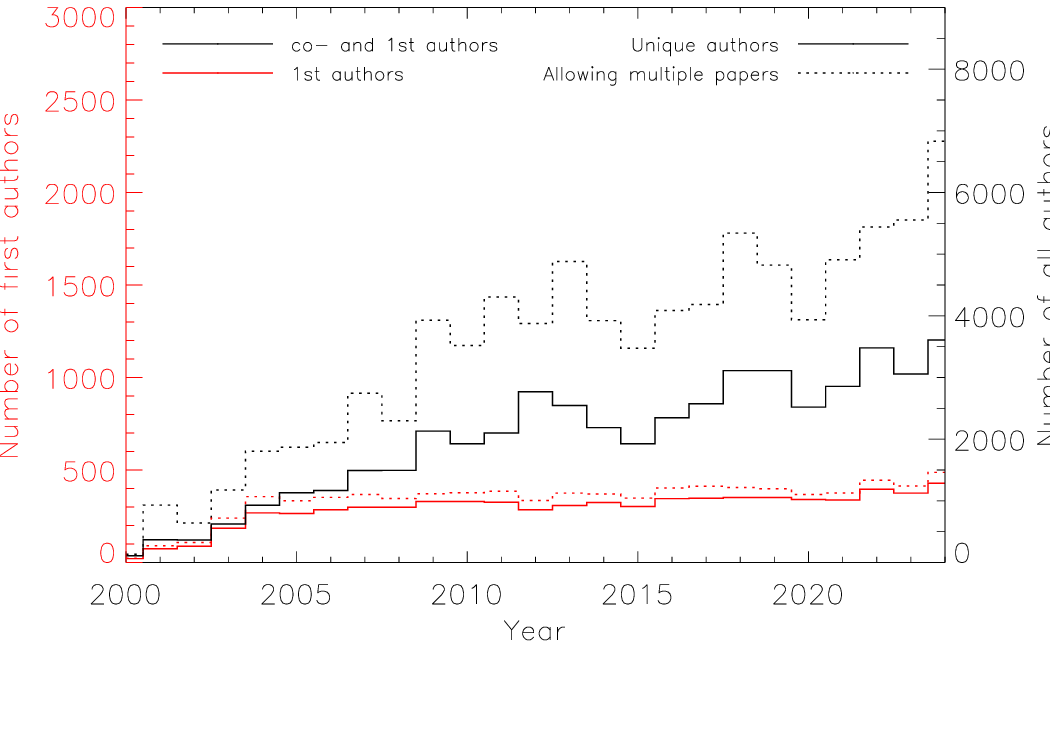}}
\caption{\label{fig:aperyear}\small Number of different authors per year (solid lines) compared
to number of all authors, thus including authors publishing multiple times in the same year
(dotted lines). Red colours represent $1^{st}$authors (values in left y-axis) while black
colours represent all authors (values in right y-axis). It can be seen that few people managed
to write more than one first-author \xmm\ paper per year while contributing as co-authors
to more than one \xmm\ paper per year occurs frequently.
}
\end{figure}

In order to characterise the community working with \xmm\ data, we analyse in this
section the author statistics in terms of number of authors having published an \xmm\
paper for the first time and for the last time until the end 2024,
compared to the number of different authors per year, duration of activity and
productivity of authors, and team sizes.\\ 

In Fig.~\ref{fig:aperyear}, we present the evolution of the number of different authors using
solid histogram lines, distinguishing between $1^{st}$ authors (red, corresponding to the
red y-axis) and all authors ($1^{st}$ authors plus co-authors, black with the black y-axis).
Additionally, we show the total number of authors, including cases where the same individual
has published multiple papers in a single year. The figure illustrates that very few individuals
have published more than one $1^{st}$-author paper per year. Meanwhile, many scientists have
co-authored multiple \xmm\ articles within the same year.
The number of different authors increased steadily until $\sim2010$, when it stabilised at
$\sim 330$ $1^{st}$authors per year, with a possible increase in the last three years. 
For 2011 to 2020 we obtained a mean of 327 with a deviation of 24.
At the same time, team sizes continued to increase until $\sim 2013$, with typical team sizes of 3-6 authors.

\begin{figure*}[!ht]
\resizebox{\hsize}{!}{\includegraphics{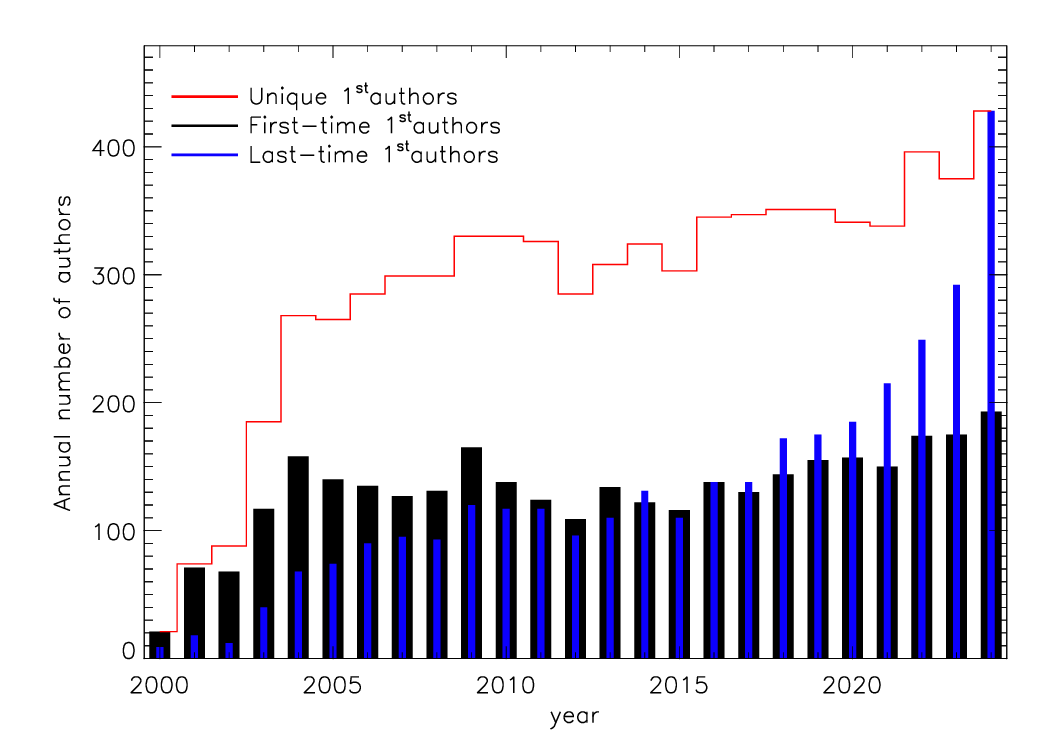}\includegraphics{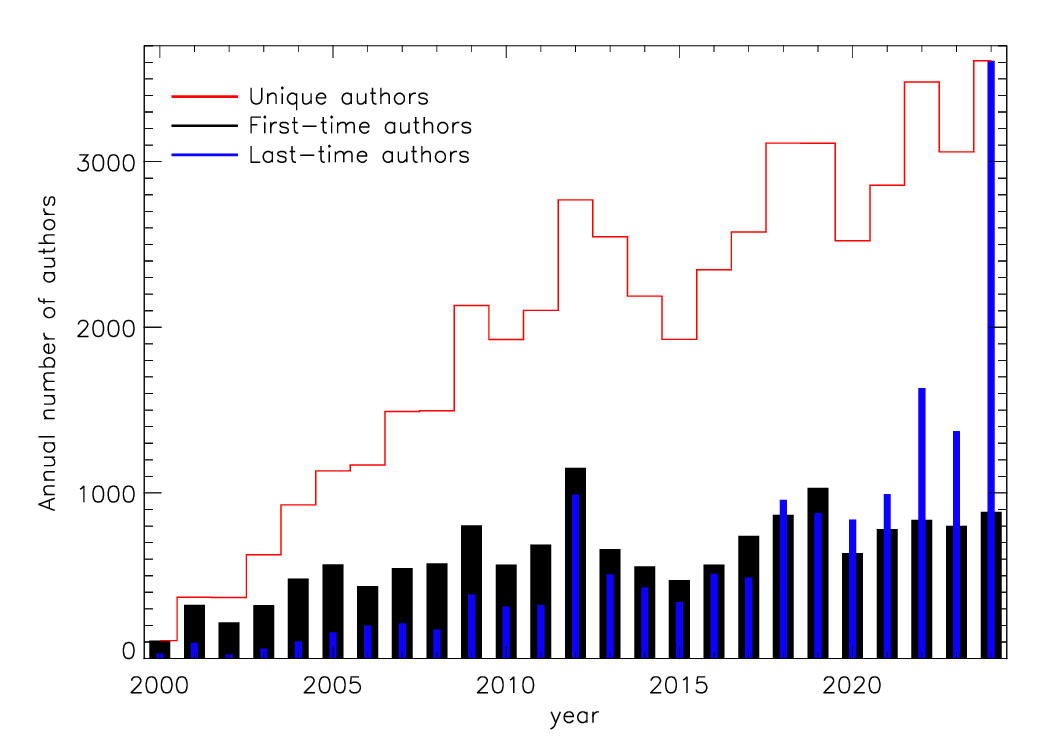}}
\caption{\label{fig:newauth}\small Evolution of the number of $1^{st}$ (left) and all (right) authors having been involved for the first time (black histogram bars) and for the last time (blue histogram bars) as authors in XMM-Newton articles. The solid red histogram lines show, for reference, the evolution of number of unique authors (same as the respective solid red/black histogram lines in Fig.~\ref{fig:aperyear}).
}
\end{figure*}

In Fig.~\ref{fig:newauth} we show the evolution of first-time authors (black histogram bars)
and last-time authors as of 2024 (blue histogram bars) focusing on $1^{st}$authors in the left panel
and all authors in the right panel. From 2004 to 2019 we found an average of 134 first-time authors per
year with a
dispersion of 15, a similar trend has already been reported by \cite{xmmpubs}. This trend has continued
and even increased after $\sim2019$. 
The number of departing $1^{st}$authors increased until 2010, and since then has remained
stable at about the same rate as the number of new authors.
The fact that the number of new $1^{st}$authors (black) is equal to the number
of departing $1^{st}$authors (blue) since 2010 implies a constant number of
$1^{st}$authors since then.

The number of last-time authors (blue bars) towards the end of the examined time frame
is of interest to estimate typical time scales on which new articles are published.
It is evident that in the final time bin, 2024, the number of distinct authors (red line)
is equivalent to the number of authors who have ceased publishing (blue bars). 
As not all authors publish annually, the rate of last-time authors will increase
gradually, with the growth rate indicating typical time scales on which new articles are
published.

As illustrated in the left panel of Fig.~\ref{fig:newauth}, representing $1^{st}$authors,
the divergence from the consistent rate of departing authors is observed from around 2017
onwards. In contrast, for all authors (right panel), the increase is considerably more
pronounced, beginning around 2022.
From this, we can presume that the publication of articles written by $1^{st}$authors
can take up to six years, while participation in articles co-authored by first-time and last-time authors occurs less than every three years, though this is speculation only based on available metrics.

From the divergence of departing authors, we derive a constant community since 2010
with approximately 4300 first-time authors of which 570 are active lead authors.

\begin{figure*}[!ht]
\resizebox{\hsize}{!}{\includegraphics{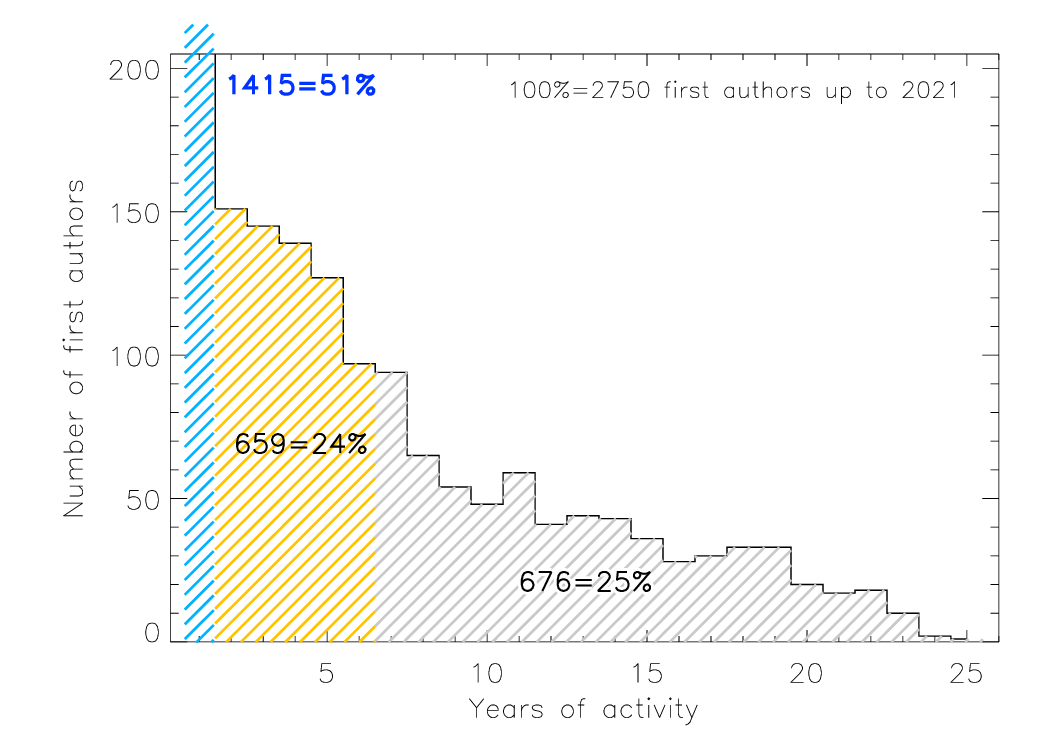}\includegraphics{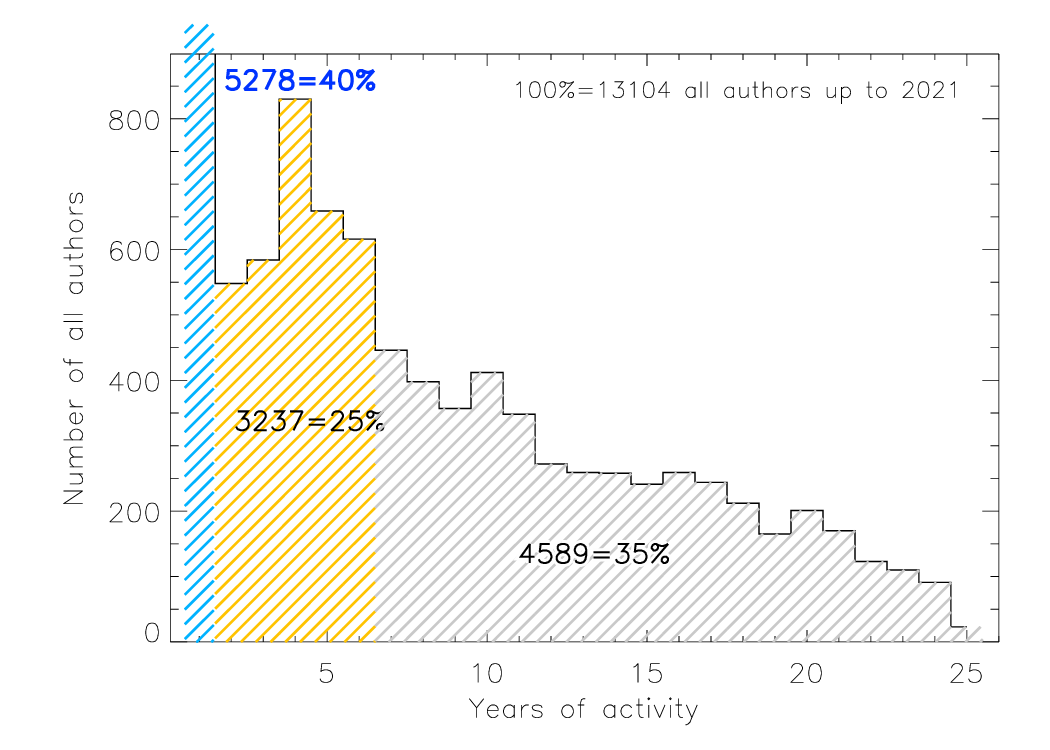}}
\caption{\label{fig:alife}\small Years of activity of first authors (left) and all authors (right), calculated from the number of years between latest and first \xmm\ article (including first year). The colours indicate the number of respective authors having been active for 1 year (light blue, exceeding the graph to the respective values of 1415 and 5278), for 2-6 years (orange) and more than 6 years (grey). 542 $1^{st}$ and 2523 $1^{st}$+co-authors with a first publication after 2021 are excluded. 
}
\end{figure*}

The number of years between the publication dates of the first and last paper for each
author (including first year), a measure of the duration of activity with \xmm\ data,
is shown for $1^{st}$ and all authors in Fig.~\ref{fig:alife}. 
To mitigate the cut-off effect, 524 $1^{st}$ and 2523 $1^{st}$+co-authors with a
first publication after 2021 are excluded.
Both curves contain a sharp peak at 1 year of activity, i.e. authors who essentially
wrote only one \xmm\ paper.
These are 52\% of $1^{st}$authors (left) and 40\% of all authors (right).
24\% of $1^{st}$authors and 25\% of $1^{st}$+co-authors have been active for 2-6 years and
the rest have published for more than 6 years. The latter figures (676 and 4589)
are consistent with the estimates of the absolute size of the core community derived
from Fig.~\ref{fig:newauth}. The fractions of articles published by each of the three
groups of 1-year, 2-6 years and $>6$ years of activity are 19\%, 24\%, and 57\%, respectively.\\

The somewhat broader peak between 2-6 years of activity for all authors (right) could
be related to the typical career progression after entering research, consisting of Masters
(1 year) and PhD projects (3-6 years) plus some initial time as an early postdoc.
Researchers may work with \xmm\ data during their early careers and then either leave academic
life or move on to other areas of research.\\

\begin{figure}[!ht]
\resizebox{\hsize}{!}{\includegraphics{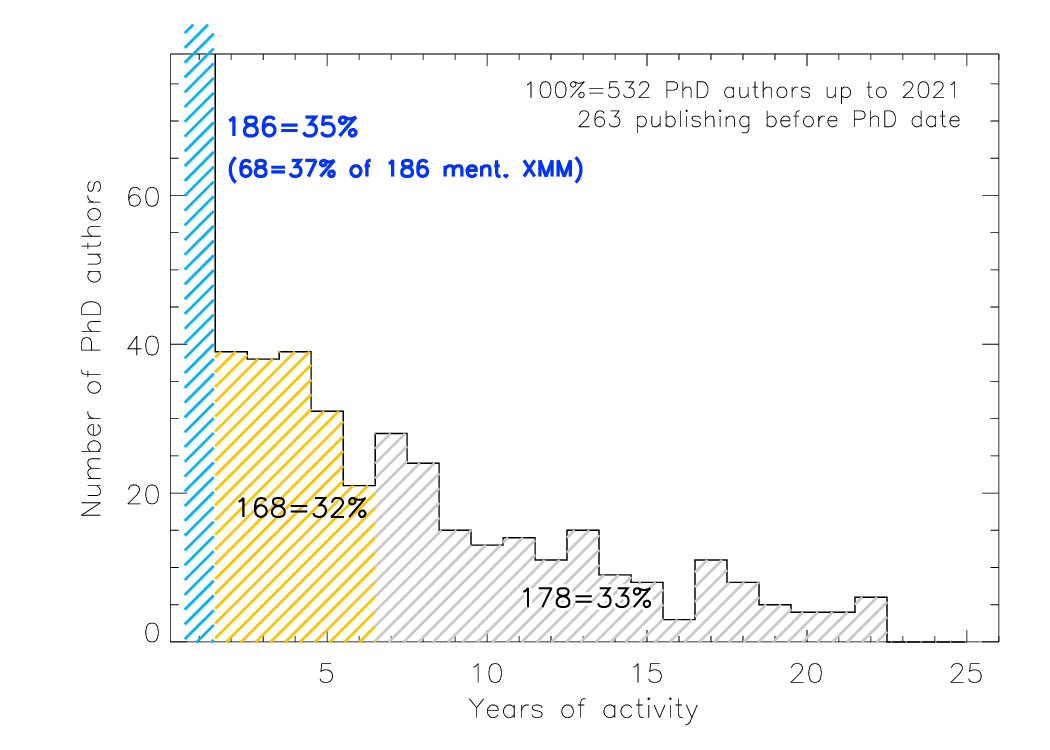}}
\caption{\label{fig:alifePhD}\small Same as Fig.~\ref{fig:alife} for 532 known PhD theses published
before 2021 (see text).
}
\end{figure}

To verify this hypothesis, we analysed the author statistics of researchers
having obtained their PhD degree based both on scientific results making use of \xmm\
observations and numerical predictions or on "technical" work (hardware and software
development, calibration, or operations). The \xmm\ project has compiled the
information\footnote{https://www.cosmos.esa.int/web/xmm-newton/phd-theses}
of 541 PhD theses with year of PhD theses, author name, thesis title, name of
supervisor, institute, country and (if available) the ADS bibcode. We focus on PhD theses
completed during the life time of \xmm\ and thus exclude one thesis from 1991, thus a
sample of 540 theses. The \xmm\ project has not done an active investigation (as for refereed
publications), rather scientists are encouraged to inform about PhD thesis making use of
XMM-Newton data via invitations in the Newsletter with approximately yearly frequency up to
mid 2023. This number must thus be treated as a lower limit as there is no
systematic process that would guarantee all theses to be ingested into the \xmm\
PhD database. We applied the same streamlining of author names as for our main
sample to search for all publications by PhD authors and generated the same type
of graphical illustration of years of activity as shown with Fig.~\ref{fig:alife}.
The result is shown in Fig.~\ref{fig:alifePhD} for 532 theses published before 2021
where it can be seen that the fraction of PhD authors having published \xmm\ related
articles for only one year is much lower (35\%) compared to the entire sample
(51\%) while the fraction of 2-6 years activity is higher at 32\% compared to
24\% for the entire sample. In fact, 25\% of articles
in the 2-6 year peak are PhD authors while only 13\% of the one-year activity are
\xmm\ PhD authors.
Out of all 540 PhD authors (including those publishing their thesis after 2021)
we identified 264 (more than half) who have already published
before publishing their PhD thesis. This supports our hypothesis that the 2-6 year
peak consists partly of early-career scientists.\\

\begin{figure}[!ht]
\resizebox{\hsize}{!}{\includegraphics{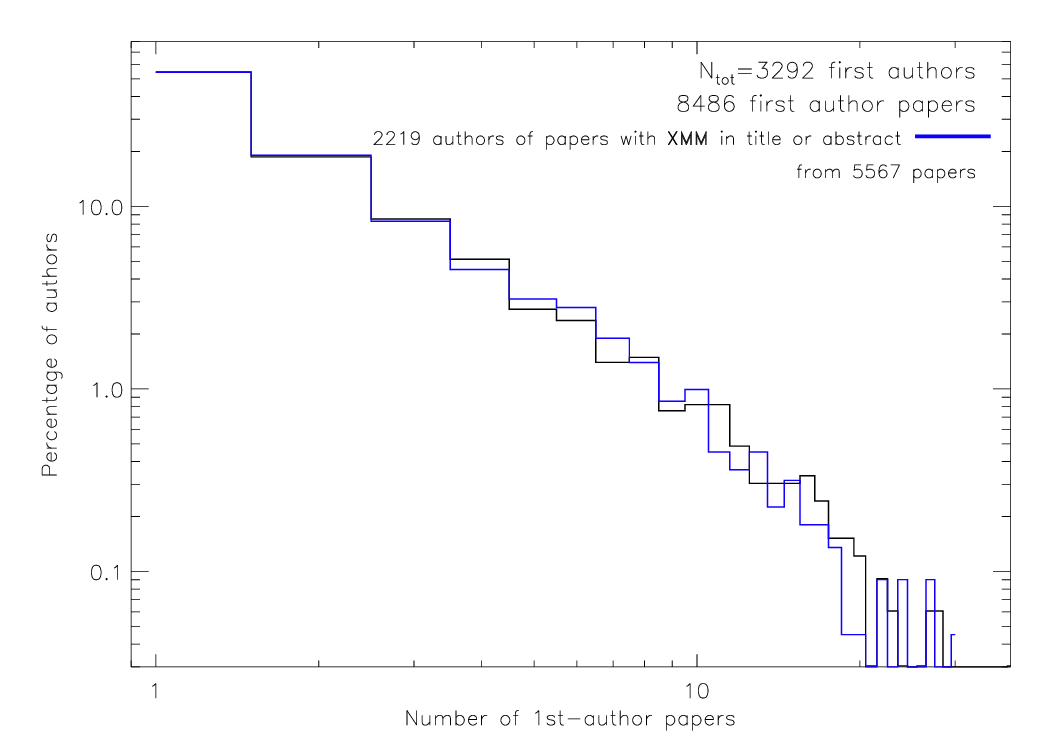}}
\caption{\label{fig:npaps}\small The number of \xmm\ Class I articles per $1^{st}$author as an indicator of productivity.
}
\end{figure}

As an indicator of productivity, we counted for each $1^{st}$author the respective total number of
$1^{st}$author papers, showing in double-logarithmic units in Fig.~\ref{fig:npaps} the fraction of
authors for each productivity bin.
As concluded from Fig.~\ref{fig:alife}, many authors have published a single \xmm\ paper during the
reference period 2000-2024 (thus 25 years) while there is a highly productive core community of authors having
published during the entire life
time of \xmm. It can also be seen that hardly any author has published more than
25 $1^{st}$author papers during the 25-year reference period, consistent with the
conclusion from Fig.~\ref{fig:aperyear} that only a small percentage of authors
manage to publish more than one $1^{st}$author \xmm\ paper per year.\\

\begin{figure}[!ht]
\resizebox{\hsize}{!}{\includegraphics{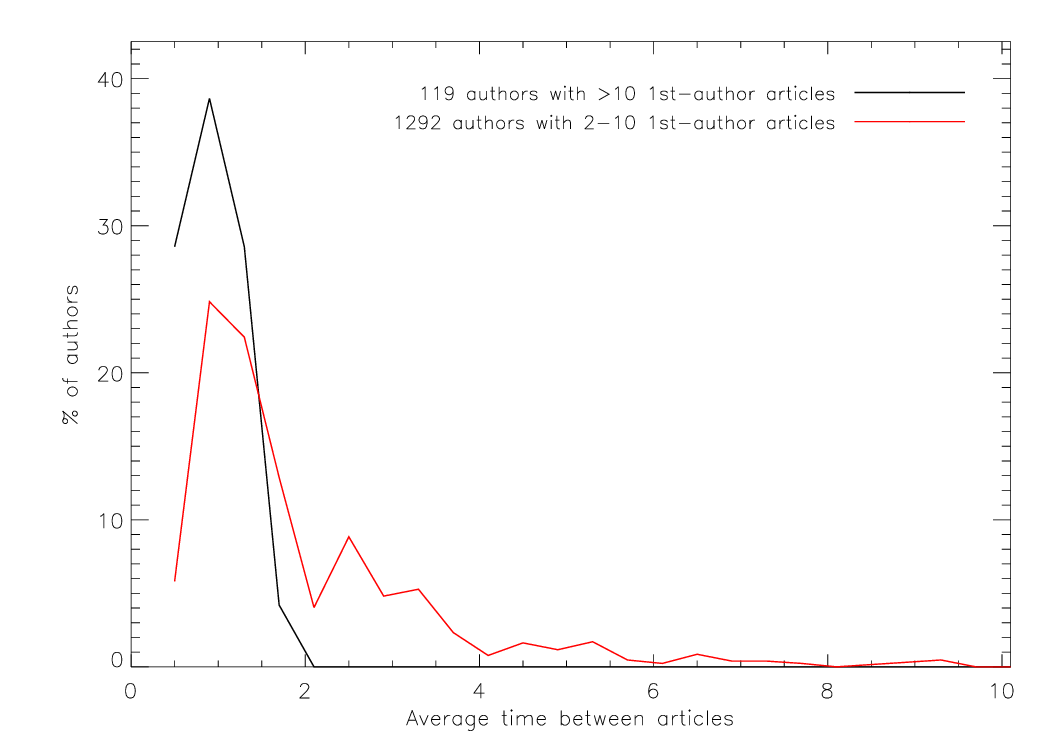}}
\caption{\label{fig:prate}\small Distribution of average elapsed time (in years) between articles for authors
publishing longer than 10 years (black) and between 2-10 years (red).
}
\end{figure}

Another indicator of productivity is presented with Fig.~\ref{fig:prate} where the distribution
of average time between articles is shown, calculated for each $1^{st}$author as the ratio of the total
number of articles (Fig.~\ref{fig:npaps}) by the total duration of activity (Fig.~\ref{fig:alife}).
Authors having only published one article are excluded, and two distributions
are shown for long-term authors publishing more than 10 years and short-term authors publishing
between 2 and 10 years.
Fig.~\ref{fig:prate} shows that there is a highly productive community of
119 authors who have published an \xmm\ article at least every two years
for more than 10 years.
 
\subsection{Usage Statistics}
\label{sect:usestat}

The first observation considered for this study started on 2000-01-19T15:29:24.0 (in revolution 21)
and the last one ended on 2024-12-31T10:15:48.0 (revolution 4590). 16894 observations with at least
one science exposure were considered for this study which combine to a total of 556\,Ms of
available observing time (thus on average 33\,ks per observation).
In the current version of the XSA, the total observing time appears to add up to 609 Ms. However,
for some mosaic programmes\footnote{Mosaic programmes are a series of nearby, usually slightly overlapping, pointings with the purpose to cover a larger sky area.}, the individual mosaic pointings are all listed with the same start/stop times and exposure duration. We identified 516 such cases in the XSA. In these cases, we used, as the exposure time for each individual pointing, the longest individual instrument exposure within that pointing. As a result, the corrected total available observing time is 556\,Ms.\\

14176 observations have been used in at least one publication (84\% of all observations)
combining to a total of 484.2\,Ms of observing time (87\% of the total available observing time of 556\,Ms).
Several early mission performance verification observations and certain types of calibration observations, such as closed filter observations of the EPICs or source offset observations, contain limited scientific data and have therefore not been used for scientific publications.\\

In this section we have a closer look at the evolution of the fraction of data used
for scientific research, including multiple use, how quickly after observation end the first
and second article was published, and finally how the observations taken within the various
science categories were used.\\

Each \xmm\ observation has been taken for a specific purpose selected by the Observation
Time Allocation Committee (OTAC).
We have no record of the extent to which publications serve the primary purpose
of the observations they use, whether they focus on serendipitous science, or whether they
perform surveys or create catalogues making use of all observations with enough sensitivity.
To distinguish between focused articles and survey/catalogue articles, we derived from the
distribution of number of observations that 95\% of all articles use less than 48 observations
and 99\% of articles use less than 238 observations.
We declare the 1\% of articles using more than 238 observation as
survey/catalogue papers. We excluded from this list nine articles using the observations of the
LSS survey. While this is a survey, these observations were taken for the survey purpose described
in the observing proposal, and the articles publishing
these observations thus serve the purpose of their observations. In total, we flagged
84 articles as catalogue papers whereas we refer to all other articles as 'dedicated articles'.\\

\begin{figure*}[!ht]
\resizebox{\hsize}{!}{\includegraphics{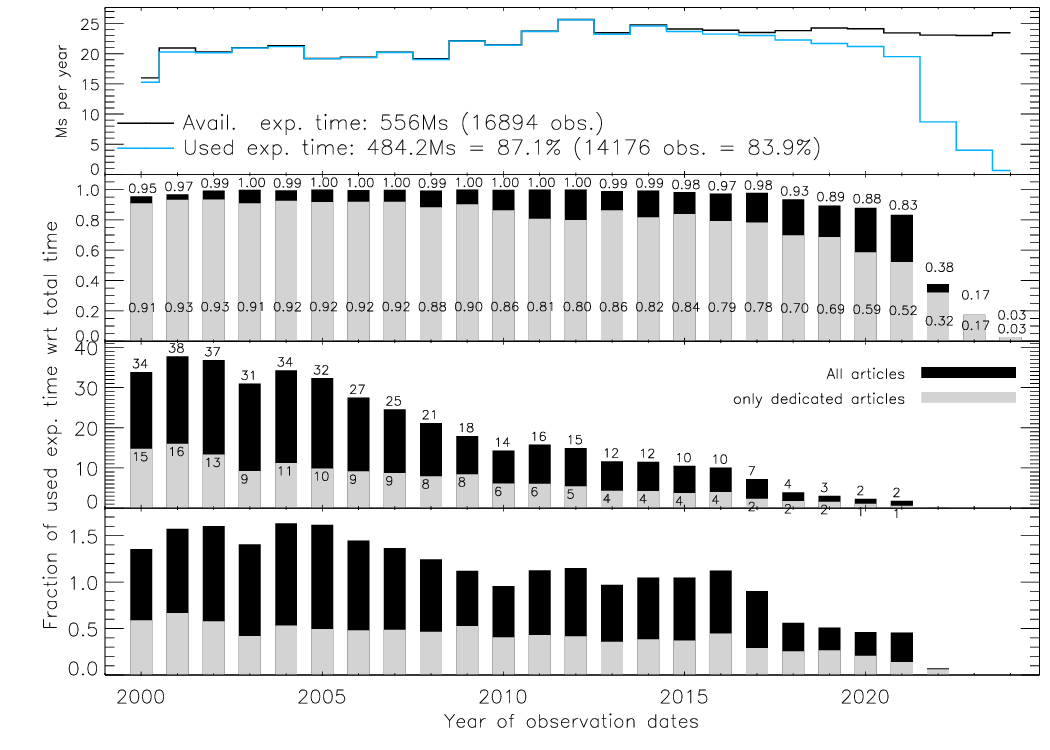}}
\caption{\label{fig:nuse}\small The top panel shows the total available science observing time taken per year
in black and the respective exposure time used in at least on publication in blue. The second panel shows the fraction of science time used in at least one publication, whereas the third panel includes multiple use by multiplying each exposure duration by the number of articles in which it was used. The bottom panel corrects for the time observations were available to be used: since older observations had more chance to be used multiple times, the numbers from the third panel were divided by the number of years before 2024. For the black bars, all articles were used while the grey bars show the same numbers excluding the 1\% of papers that we do not consider as 'dedicated articles' (see text). Multiple use naturally increases with the age of the data (third panel), and correcting for this effect yields a roughly constant recycling rate (bottom).
}
\end{figure*}

The evolution of usage of \xmm\ observations is illustrated in Fig.~\ref{fig:nuse}.
As a reference, in the top panel the amount of available exposure time is shown for each year
while the panels below illustrate the fraction of these time budgets that have been used in publications
at least once (second panel), multiple times (third panel), and multiple times normalised by the age of the
data, thus how long they were available for publications (bottom panel). The light grey bars represent the usage
limiting the sample to the 99\% of articles flagged as dedicated papers (see above) while the black bars
represent the outcome from all articles.\\

The fractions in the second panel were computed by adding up all exposure times of observations that were
used in at least one publication. Until $\sim2009$, the usage in dedicated papers ranges above 90\% while
from 2010 until
$\sim2016$ above or around 80\%. In other graphics, e.g. Fig.~\ref{fig:newauth} the effects of the
cut-off at 2024 can be seen to reach back some 5-6 years
which implies publishing the results from \xmm\ data takes several years. We thus expect the usage fraction
to also increase for observations taken after 2017.\\

The fraction of multiple-use papers in the
third panel were calculated by adding up the exposure times of observations multiplied by the number of
articles in which they were used. Including large survey papers pushes reuse to a factor of 30 for the
earliest observations while the trend for dedicated papers is more stable, around a factor of 10. Since data in
the archive for a longer time have a higher chance of being used in publications, we show in the bottom
panel the statistics from the third panel divided by the number of years before 2024.

\begin{figure*}[!ht]
\resizebox{\hsize}{!}{\includegraphics{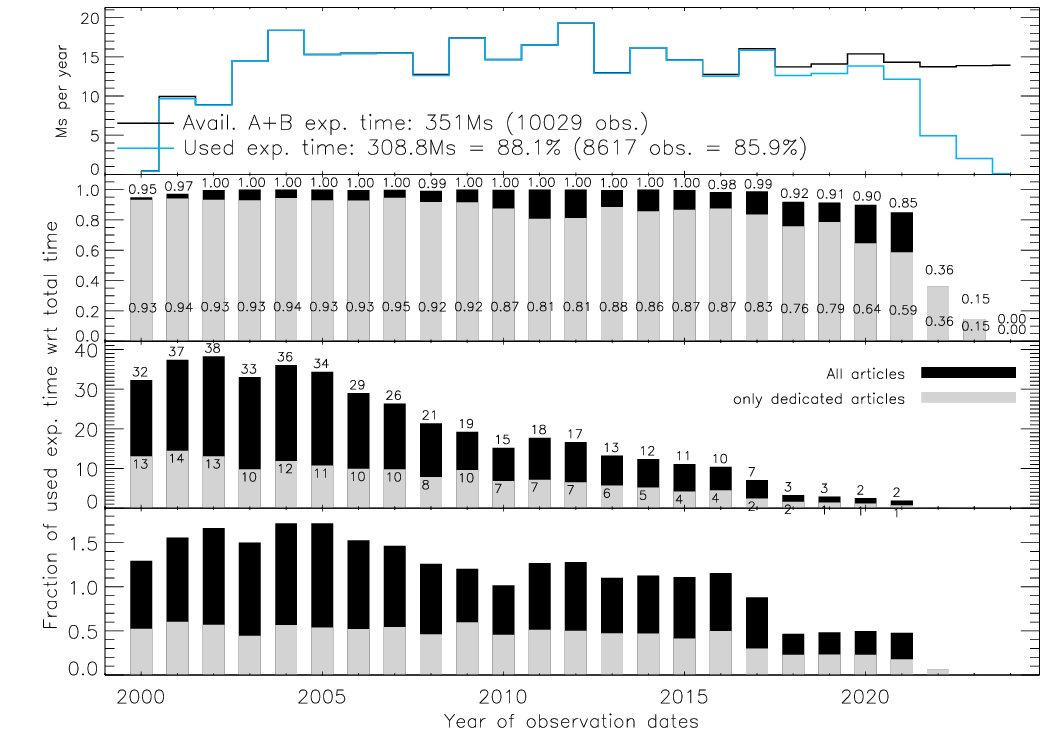}\includegraphics{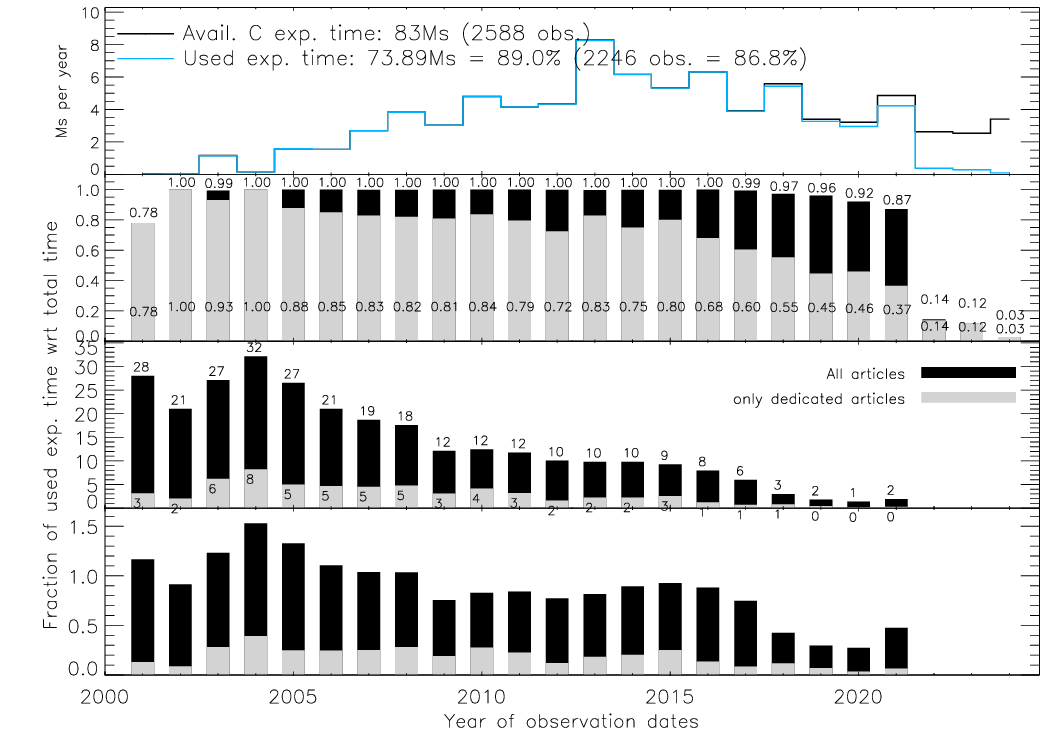}}
\caption{\label{fig:nuse_ABC}\small Same usage statistics as Fig.~\ref{fig:nuse} for only A+B-priority observations
(left) and C-priority observations (right). The statistics of using C-priority observations at least once is about the same as for A+B targets indicating that proposing teams of C-priority observations are equally interested in these data. However, multiple use in non-catalogue papers is significantly lower for C priority observations.
}
\end{figure*}

The general conclusion from Fig.~\ref{fig:nuse} is that the usage of \xmm\ data is extremely high
and stable over the years. In Fig.~\ref{fig:nuse_ABC}, we compare the same statistics for A+B
priority targets with C priority targets.
C targets have been rated lower during the OTAC review
but are included in the observing programme of an observing cycle as 'filler' targets, thus
are only observed at times when no suitable A or B target is available for scheduling.
As can be observed in the respective top panels, the observing time in the C category was initially very low. This was due to an overallocation of observing time for the guaranteed time programme and AO1, as the impact of radiation was not known before launch.
The proportion of C-priority observations employed on at least one occasion is substantial. However, the recycling rate by dedicated articles is lower than for A and B-priority observations.\\

\begin{figure*}[!ht]
\resizebox{\hsize}{!}{\includegraphics{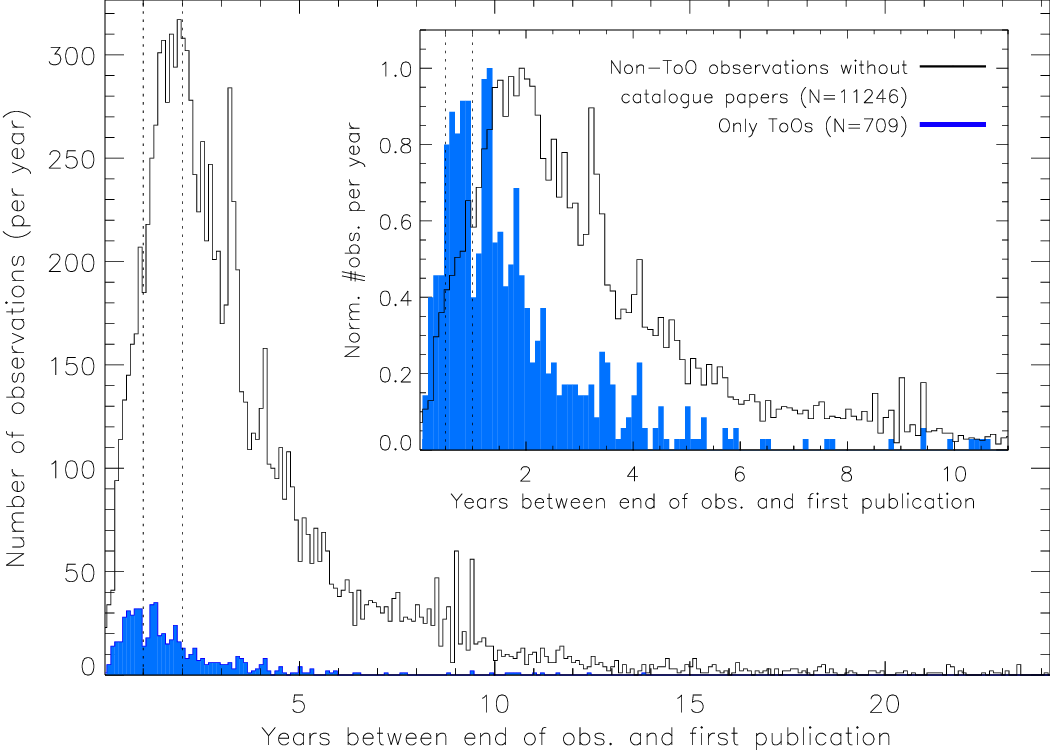}\includegraphics{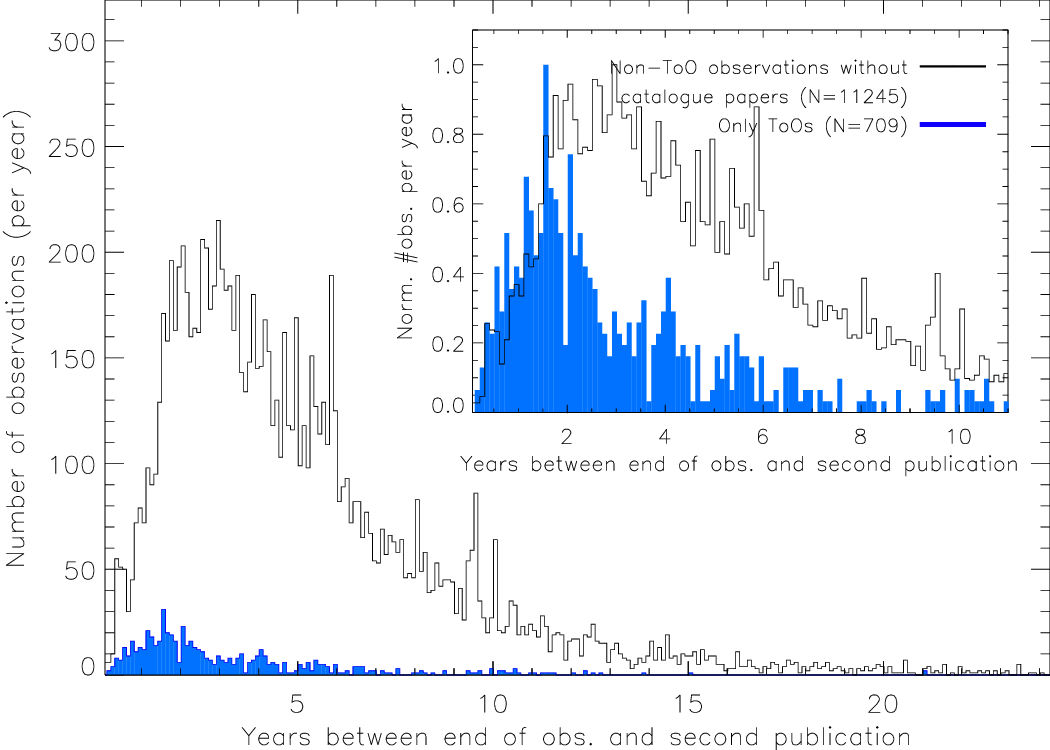}}
\caption{\label{fig:firstpubs}\small Elapsed time between the date when observations were taken and when the first (left) and second (right, on same vertical scale) refereed article was published. Only 'dedicated' publications are considered for these graphs. The blue shaded area shows the same distribution only for ToO (DDT) observations. The inset shows the same distributions in normalised form, and clearly, ToO observations are published faster than regular observations (note that the black histograms include the ToO observations). The vertical dotted lines mark the proprietary period and twice this amount for GO observations (1 year) in the main panel and for ToOs (6 months) in the inset.
}
\end{figure*}

Fig.~\ref{fig:firstpubs} illustrates how long it typically takes to publish \xmm\ data. This question
is of interest to understand the priority \xmm\ data have on the agenda of the community and possibly
the complexity of the data analysis. These questions can only be answered by focusing on the 'dedicated'
publications as survey papers use all available data at the time of processing, irrespectively how old
they are. We thus exclude the 84 survey papers for construction of the distribution of first and
second publication times after end of observations. The first-publication distribution peaks around 1.5-2 years,
a bit longer than the proprietary period of Guest-Observer (GO) observations (1 year, this, and twice that,
are marked with the vertical dotted lines in the main panel). This indicates that the proposers holding data rights
attempt to publish their data before they become publicly available but don't quite achieve that. Since
competitors getting access to the data once the proprietary period has expired will also need some time
to obtain and publish results, proposers may not be too concerned to publish by the day the data become public,
although a small peak is seen at exactly one year. Indeed, a narrow secondary peak is seen at
$\sim3.5$ years, 1 year after the main peak. It may be coincidence that the time between these two peaks
is the same as the proprietary period, but if real it can be interpreted as scientists picking up
unpublished data when they become available take about the same $\sim1-2$ years as proposers to publish them.
This result indicates that \xmm\ data enjoy a high priority on the agenda of the scientific community.\\

The blue histogram focuses on 709\footnote{https://www.cosmos.esa.int/web/xmm-newton/too-details
https://www.cosmos.esa.int/web/xmm-newton/too-details-archive
https://nxsa.esac.esa.int/nxsa-web/} unanticipated Target of Opportunity (ToO) observations and
observations approved under the Director's Discretionary Time (DDT) budget until the end of 2024 that
have been used in at least one publication. These observations were proposed by community scientists
and, if approved, the data became immediately public, or, in most of the cases,
the proposers were awarded 6 months of proprietary data rights (vertical dotted lines
in inset mark these 6 months and twice of that), thus have somewhat
higher pressure if they wish to publish the results free of competition. The comparison of the normalised
histograms in the inset shows that these ToO observations are typically published faster than regular
GO observations; possibly also with a secondary peak one year later, i.e., scientists picking up unpublished
data from ToO observations need about the same time as scientists analysing GO data. 
The accelerated pace of publishing ToO observations may be indicative of a tendency to expedite the dissemination of novel findings in order to gain a competitive advantage over observations obtained through alternative facilities.\\

The distributions of second publications after respective end of observation (right panel in
Fig.~\ref{fig:firstpubs}) is obviously shifted further to the right and is broader reflecting that
once the first article is published, the time pressure is lower. The absolute numbers are amazingly
high, supporting the conclusion of a high re-usage of \xmm\ observations.\\

\begin{figure}[!ht]
\resizebox{\hsize}{!}{\includegraphics{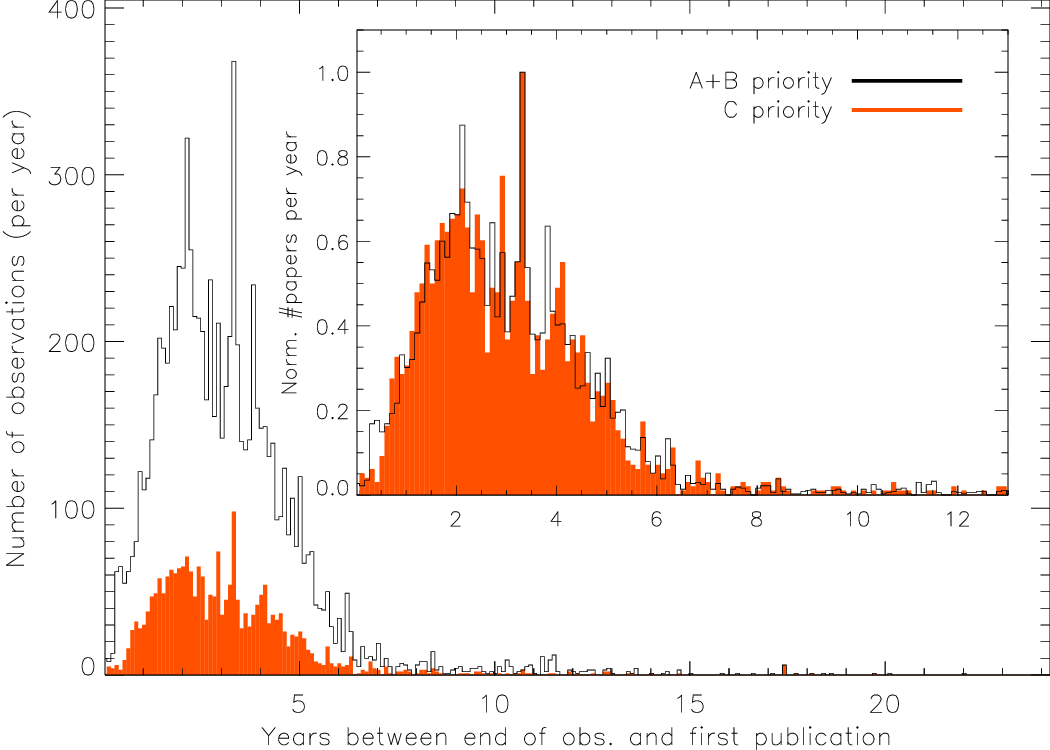}}
\caption{\label{fig:firstpubs_ABC}\small Same as left plot in Fig.~\ref{fig:firstpubs} focusing on A+B and for C proposals as indicated in the legend. The C-priority observations are published at roughly the same speed as A+B observations.
}
\end{figure}

In Fig.~\ref{fig:firstpubs_ABC}, the same distributions are compared for high-priority A+B GO
observations and lower-priority C observations, and the normalised distributions match demonstrating
that C-priority observations are published equally fast as high-priority A+B observations. The
same conclusion can be drawn from comparing the distributions of A+B and C observations being
published for the second time after end of observation which are not shown here.\\

\begin{table}
\fontsize{9pt}{9pt}\selectfont
  \caption{\label{tab:categories}\small For 14 science categories, based on the HEASARC Class codes (2nd columns), the total number of observations (third column), total observing time (fourth column), and number of publications (last column).}
  \begin{tabular}{p{3.4cm}rp{.6cm}p{.6cm}p{.6cm}}
  \hline
\multicolumn{2}{l}{Science Category \hfill Codes$^a$} & \#obs$^b$ & Ms$^c$ & \#Pubs\\
  \hline
$(1)$ Solar System objects & 8000-8999 & {\bf 265} & {\bf 7.23} & {\bf 1738} \\
$(2)$ Extrasolar Planets, Brown dwarfs, Protostars & 1850-1899 & {\bf 103} & {\bf 2.67} & {\bf 1459} \\
$(3)$ Stars/WDs &  & {\bf 2111} & {\bf 58.54} & {\bf 20646} \\
 \ \ \ \ -- Stars & 1900-2999 & 1897 & 54.14 & 19440 \\
 \ \ \ \ -- White dwarfs & 4000-4999 & 214 & 4.40 & 1206 \\
$(4)$ Supernovae and Hypernovae & 9100-9299 & {\bf 81} & {\bf 2.68} & {\bf 688} \\
$(5)$ SNR, Nebulae, Diffuse emission & 3000-3999 & {\bf 1425} & {\bf 48.00} & {\bf 18341} \\
$(6)$ Binaries &  & {\bf 1911} & {\bf 66.29} & {\bf 21556} \\
 \ \ \ \ -- Cataclysmic Variables & 1600-1699 & 513 & 13.17 & 7139 \\
 \ \ \ \ -- X-ray binaries, QPO & 1000-1599 & 1145 & 39.25 & 12113 \\
 \ \ \ \ -- ULX & 9300-9399 & 253 & 13.87 & 2304 \\
$(7)$ Gamma ray sources & 1700-1799 & {\bf 274} & {\bf 11.61} & {\bf 3225} \\
$(8)$ Pulsar, Neutron stars & 1800-1849 & {\bf 806} & {\bf 30.56} & {\bf 10351} \\
$(9)$ Galaxies & 6000-6999 & {\bf 1090} & {\bf 38.93} & {\bf 20822} \\
$(10)$ Active Galaxies &  & {\bf 3846} & {\bf 146.41} & {\bf 62140} \\
 \ \ \ \ -- AGN, Seyferts & 7000-7199 & 1741 & 65.78 & 28822 \\
 \ \ \ \ -- AGN (Liners) & 7400-7499 & 34 & 1.11 & 1048 \\
 \ \ \ \ -- Radio/IR Galaxies & 7600-7999 & 344 & 16.57 & 7215 \\
 \ \ \ \ -- QSO & 7200-7299 & 941 & 34.64 & 14412 \\
 \ \ \ \ -- BL Lac & 7300-7399 & 406 & 14.92 & 5296 \\
 \ \ \ \ -- Radio galaxies & 7500-7599 & 235 & 8.35 & 4090 \\
 \ \ \ \ -- Supermassive black holes & 9400-9499 & 145 & 5.05 & 1257 \\
$(11)$ Groups of Galaxies & 5100-5299 & {\bf 391} & {\bf 13.18} & {\bf 5061} \\
$(12)$ Clusters of Galaxies & 5000-5099 & {\bf 2023} & {\bf 67.33} & {\bf 33585} \\
$(13)$ X-ray background & 5500-5599 & {\bf 265} & {\bf 5.08} & {\bf 7017} \\
\multicolumn{2}{l}{$(14)$ Cosmology, Deep/Large Fields$^d$} & {\bf 1398} & {\bf 34.98} & {\bf 24742} \\
&Total:&{\bf 15989}&{\bf 533.50}& \\

  \end{tabular}
$^a$http://heasarc.gsfc.nasa.gov/W3Browse/class\_help.html\\
$^b$Number of observations containing at least one science exposure\\
$^c$Total exposure time in $10^6$\,seconds (Ms)\\
$^d$All observations with target names containing Marano, Lockman, CDF, CDFS, CDS, AXAF, HDF, COSMOS, MLS, SZE, Stripe, Deep, XXL, ELAIS, LSS, EDFF, MW Plane, XBCS, XBCSM
\end{table}

\begin{figure}[!ht]
\resizebox{\hsize}{!}{\includegraphics{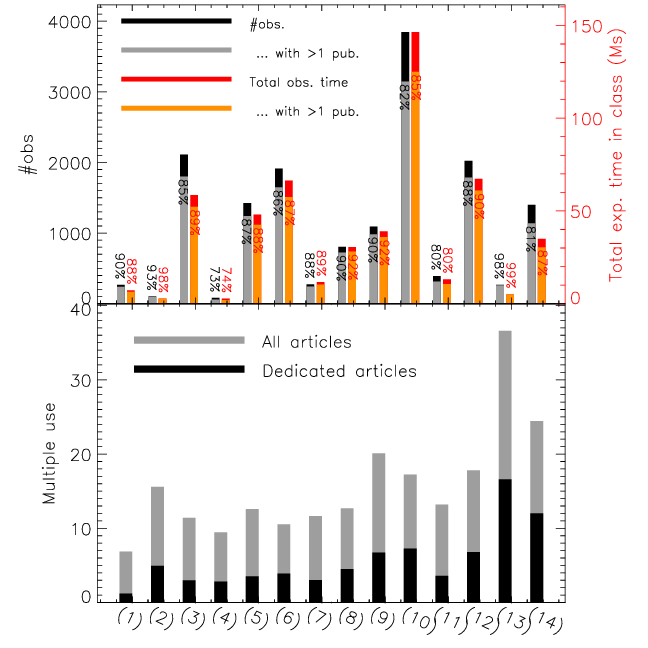}}
\caption{\label{fig:ttypes}\small Graphical representation of usage statistics per science category (see Table~\ref{tab:categories}). {\bf Top panel}: 'Investment versus return' in terms of number of observations (black/grey) and exposure time (red/orange) with grey/orange indicating the respective numbers/exposure durations used in at least one publication. The fractions of observations and total exposure time used in at least one publication are marked with percentages in each bar for each science category. {\bf Bottom panel}, multiple use: product of observing time and respective number of publications.
}
\end{figure}

With Table~\ref{tab:categories} and Fig.~\ref{fig:ttypes} we focus on the science categories. \xmm\ is
of interest to a large range of research fields, and we defined 14 science categories for the assessment
of the productivity of \xmm\ observations in each of them. For most (15989) observations, the XSA contains
a column with a valid HEASARC Class code (code 9999=unidentified excluded) provided by the respective
Principal Investigators (PIs) during proposal submission. These codes thus subjectively reflect the
science goals of the proposals without necessarily constituting a unique target classification. For example,
an observation of an F star with an exoplanet may either be classed with 2400 (the F star) or 1850 (Extrasolar
planet). The 14 science categories combine ranges of HEASARC codes as given 
in the second column of Table~\ref{tab:categories}. For the last category of
Cosmology and Deep fields, we included all observations (regardless of the value of class code)
with target names including the large extragalactic fields designations quoted in the table notes.

In columns three and four, we list the number of observations and total amount of exposure duration
(thus the investment from the \xmm\ project), and in the last column the number of publications
(thus the science return from this investment). Fig.~\ref{fig:ttypes} gives a graphical
illustration of the investment versus return (top panel) and the usage in terms of fraction of single use
and multiple use (bottom panel).
One can see that the largest investment was made for category 10, Active Galaxies, and the smallest for categories 2 and 4, Extrasolar Planets, Brown dwarfs \& Protostars and Supernovae \& Hypernovae, respectively.
The absolute number of unused observations and respective unused observing
time are indicated by dark black and red bars on top, respectively, whereas the respective fractions of
invested time are given at each bar. All science categories are close to the average of
80\% observing time being used at least once, with the category of (4) Supernovae and Hypernovae only reaching
$\sim70$\%; note that this is the category with the lowest number of observations.
Multiuse (bottom panel) is particularly high for X-ray background observations which is understandable since
research dealing with X-ray background will always seek as many observations as possible.
If excluding survey/catalogue papers (i.e., dedicated articles, black bars), the lowest multiuse can be seen
for the Solar System category which can be explained by the higher complexity of
data analysis (moving objects), the smaller community and weak detections.

\section{Summary and Conclusions}

During the reference period 2000-2024, a total of 8486 refereed articles related to \xmm\ have been published,
attracted by a total of 16894 available observations with at least one science exposure comprising a total
time budget of 556\,Ms. The publication rate is stable at $\sim 400$ publications
per year, where the fraction of articles making use of the primary \xmm\ catalogues has grown in recent years.\\

We present detailed statistics, and found the following results characterising the \xmm\ community
\begin{itemize}
\item Mentioning the mission name in the title or abstract can be considered a necessary, though not
sufficient, indicator of the relevance of mission data to a study’s conclusions. Until around 2008,
XMM-Newton was mentioned in the title in 30-60\% of all articles, providing an approximate lower limit
on the rate of articles for which XMM-Newton data was essential. After 2008, this proportion stabilised at
approximately 20\%.

\item Meanwhile, XMM-Newton was mentioned in the abstract or title in about 60\% of articles at all times,
providing a lower limit on the rate of articles for which XMM-Newton data was relevant to the conclusions.
From the beginning of the mission until around 2008, this proportion gradually declined from 100\% to 80\%.

  \item There are 15627 different scientists involved in \xmm\ publications, of which 3292 different scientists have published at least one \xmm\ article as $1^{st}$author.

   \item Each year, $~130$ authors wrote a $1^{st}$author \xmm\ article for the first time. Between 2010 and 2016, the annual number of $1^{st}$authors publishing an \xmm\ paper for the last time is roughly the same (Fig.~\ref{fig:newauth}), suggesting a constant community.

   \item The size of the $1^{st}$author core community can be estimated at 570 different $1^{st}$authors, $\sim 25$\% (Figs.~\ref{fig:newauth},\ref{fig:alife}).

   \item The size of the core community including all authors can be estimated to be $\sim4300$ different scientists, $\sim35$\% (Figs.~\ref{fig:newauth},\ref{fig:alife}).

   \item About half of all 1$^{st}$authors published for only one year while a quarter were active for 2-6 years and more than 6 years, respectively (Fig.~\ref{fig:alife}). The same analysis for PhD theses published during the life time of \xmm\ results in one third of authors publishing for only one year, 2-6 years and more than 6 years, respectively (Fig.~\ref{fig:alifePhD}).  

   \item The core community (publishing more than 6 years) typically publishes less than 1 article per year as $1^{st}$authors, while publishing at least every six years. The frequency of publishing \xmm\ articles as co-authors is much higher, less than every three years (Fig.~\ref{fig:newauth}).

   \item A highly productive core community of $\sim120$ scientists who have been active for more than 10 years publish an XMM-Newton article at least every two year as $1^{st}$authors on average (Fig.~\ref{fig:prate}).

\end{itemize}

The average fraction of the total available science exposure time of 556\,Ms that was used in at least one refereed
publication is 87\%. The following results are found from detailed usage studies:

\begin{itemize}
   \item 95\% of all articles use less than 49 observations and can be considered focus articles.
   \item 1\% of articles using more than 238 observations are defined as survey/catalogue papers.
   \item Accounting for 99\% of all non-survey articles (referred to as 'dedicated papers'), the publication fraction of 80-90\% of the available observing time remained stable until 2015. The decline in usage during the last six years is explained with typical time scales for publishing \xmm\ results (Figs.~\ref{fig:nuse}, \ref{fig:newauth}).
   \item The recycling rate of \xmm\ observations ranges from factors 5-15 for data taken before 2017. If including 1\% survey/catalogue articles, the recycling rate reaches a factor up to 30 (Fig.~\ref{fig:nuse}).
   \item The usage rate of A+B and C priority observations are about the same while the recycling rate by dedicated papers is lower for C priority observations (Fig.~\ref{fig:nuse_ABC}).
   \item The time between end of observation and first publication peaks around 2 years, longer than the period of proprietary data rights. A strong secondary peak is found in the distribution one year later indicating that the analysis of unpublished data picked up by the general community takes about the same time as for the proposers holding data rights (Fig.~\ref{fig:firstpubs}).
   \item The time between end of observation and first publication for ToO observations is only one year, possibly owed to the shorter duration of proprietary data rights and novelty of results (Fig.~\ref{fig:firstpubs}).
   \item The time between end of observation and first publication is the same for A+B and C priority observations (Fig.~\ref{fig:firstpubs_ABC}).
   \item The usage fraction of observations initially taken under 14 science categories range around or above 80\% for all (Fig.~\ref{fig:ttypes}).
   \item The recycling rate is highest for observations taken under the science category X-ray background (Fig.~\ref{fig:ttypes}).
\end{itemize}

From these numbers, we conclude that in its first 25 years \xmm\ has maintained a high productivity and
remains a highly productive mission with a large fraction of its data being used at least once, plus
a high recycling rate.
Observations approved under the un-anticipated ToO programme are generally published faster than regular GO
observations. C-priority observations are used at the same level as high-priority A+B observations,
albeit with lower re-use.
We also find a well-balanced distribution of usage across 14 science categories.\\

The community consists of a healthy mix of core, long-term contributors and authors publishing for shorter periods
of time, assuring both continuity
and new ideas with fresh perspectives. The long reference period allows long-term trends to be identified, and no decline in any key
performance parameters is found. These statistics thus demonstrate that \xmm\ remains stable and has a promising future ahead.

\bibliography{jn,xmmcats}

\end{document}